\documentclass{article}
\usepackage{spconf,amsmath,graphicx,booktabs}
\usepackage{hyperref}
\usepackage{multirow}
\usepackage{arydshln}
\usepackage[table]{xcolor}

\title{Towards Blind Data Cleaning: A Case Study in Music Source Separation}

\name{
  \begin{tabular}[t]{c}
    Azalea Gui$^{*1,2}$\thanks{$^*$ Work done during an internship at Sony AI. [azalea@hydev.org]} \qquad
    Woosung Choi$^{2}$ \qquad
    Junghyun Koo$^{2}$ \qquad
    Kazuki Shimada$^{2}$ \\
    \textit{Takashi Shibuya}$^{2}$ \qquad
    \textit{Joan Serrà}$^{2}$ \qquad
    \textit{Wei-Hsiang Liao}$^{2}$ \qquad
    \textit{Yuki Mitsufuji}$^{2,3}$
  \end{tabular}
}
\address{$^{1}$ University of Toronto \qquad $^{2}$ Sony AI \qquad $^{3}$ Sony Group Corporation}

\newcommand{\Frechet}{Fr\'echet}

\begin{document}
\ninept

\maketitle

\begin{abstract}
The performance of deep learning models for music source separation heavily depends on training data quality. However, datasets are often corrupted by difficult-to-detect artifacts such as audio bleeding and label noise. Since the type and extent of contamination are typically unknown, cleaning methods targeting specific corruptions are often impractical.
This paper proposes and evaluates two distinct, noise-agnostic data cleaning methods to address this challenge. The first approach uses data attribution via unlearning to identify and filter out training samples that contribute the least to producing clean outputs. The second leverages the Fr\'echet Audio Distance to measure and remove samples that are perceptually dissimilar to a small and trusted clean reference set. On a dataset contaminated with a simulated distribution of real-world noise, our unlearning-based methods produced a cleaned dataset and a corresponding model that outperforms both the original contaminated data and the small clean reference set used for cleaning. This result closes approximately 66.7\% of the performance gap between the contaminated baseline and a model trained on the same dataset without any contamination. Unlike methods tailored for specific artifacts, our noise-agnostic approaches offer a more generic and broadly applicable solution for curating high-quality training data.
\end{abstract}

\begin{keywords}
Music Source Separation, Data Cleaning, Data Attribution, Unlearning, Audio Signal Processing
\end{keywords}

\section{Introduction}
\label{sec:intro}

While the quantity of training data is widely recognized as a key driver of performance, data quality is often overlooked.
In practice, the presence of noise negatively affects a model's performance and generalization~\cite{zha25data-centric}.
This issue is evident in music source separation, where label noise (incorrect or ambiguous instrument labels) and audio bleeding (faint residual sounds from other instruments) are common in raw recordings.
The recent Sound Demixing Challenge 2023 - Music Demixing Track (MDX23) highlighted this vulnerability, showing that bleeding and label noise can lead to a significant degradation in model performance \cite{fabbro_sound_2024}.
Participants in MDX23 employed various approaches to mitigate the impact of these artifacts, such as loss truncation \cite{kang_improved_2020, kim_sound_2023, noauthor_davidliujiafengccom_mdx2023_nodate}, using noise-robust loss \cite{goswami_iseparate-sdx_2023}, and filtering data through inference SDR scores \cite{goswami_iseparate-sdx_2023}.
This collective experience highlights an important lesson: model performance is often constrained by data quality, regardless of architectural advancements.

Crucially, unlike in a curated challenge dataset, the exact type and extent of contamination in a large real-world dataset are often unknown (i.e., blind), making targeted cleaning solutions impractical. While certain issues (e.g., label noise) can be addressed by training classification models to filter suspicious samples~\cite{koo_self-refining_2023}, other artifacts (e.g., bleeding) are far more insidious. They are difficult to detect automatically, and challenging to quantify and remove without time-consuming manual labor. Thus, such artifacts pose a significant bottleneck to creating large-scale clean datasets.

We formulate this challenge as \textit{blind data cleaning}, where we aim to clean potentially noisy training datasets without prior knowledge of the underlying corruption.
In this paper, we conduct a case study for blind data cleaning in music source separation with noisy datasets.
We define the task as follows: given a noisy dataset with $N$ samples, our goal is to clean it without any prior knowledge about the type or extent of corruption, assuming access to only limited information, such as a small amount $M$ of clean data (i.e., $M \ll N$).
A cleaning method is evaluated by training a conventional music source separation model (without any noise-robust learning methods~\cite{kang_improved_2020, noauthor_davidliujiafengccom_mdx2023_nodate, goswami_iseparate-sdx_2023, noauthor_kuielabsdx23_2025}) on the cleaned dataset and measuring the resulting model's performance.

To advance towards blind data cleaning, we propose two noise-agnostic approaches: a data-attribution-based method and a distributional-metric-based method.
The former utilizes data attribution via unlearning~\cite{ko_mirrored_2024, wang_data_2025} to identify training samples that contribute the least to producing clean outputs. The underlying philosophy is that noisy or problematic training samples are expected to have smaller contributions to learning.
Following recent unlearning-based data attribution methods that adopt the mirrored influence hypothesis~\cite{ko_mirrored_2024,wang_data_2025}, we invert the setup: instead of unlearning training samples to quantify their influence, we unlearn clean samples to estimate the influence of each training sample.
The second approach leverages a distributional metric (e.g., the \Frechet{} Audio Distance~\cite{kilgour_frechet_2019, gui_adapting_2024}) to measure the similarity between the distribution of embeddings of training samples and that of a clean reference set.
One advantage of such metrics is that they operate on distributions rather than individual samples, allowing us to identify samples with out-of-distribution embeddings, even when they have different lengths.
By design, the proposed methods are not tailored to a specific artifact or task, offering a potentially robust solution for curating high-quality data across diverse domains. Moreover, they can be applied in conjunction with existing methods.

\section{Methods}

\subsection{Data Cleaning via Unlearning-based Data Attribution}
\label{methods:unlearning}

We propose a data cleaning method based on data attribution, motivated by the hypothesis that noisy samples contribute less to the trained model.
Inspired by a recent unlearning-based data attribution method~\cite{wang_data_2025, choi2025large}, we use unlearning to attribute the influence of each training sample.
To make influence estimation efficient, we adopt the mirrored influence hypothesis~\cite{ko_mirrored_2024,wang_data_2025}, which states that train-to-test influence is highly correlated with test-to-train influence.
Accordingly, instead of unlearning $N$ training samples, we unlearn a small set of $M$ trusted clean samples and measure the resulting impact on each training sample's loss.
This approach is computationally efficient since $M \ll N$.
A low impact suggests that the training sample does not align well with the characteristics of clean data, making it a candidate for removal.

\begin{figure}[t]
    \centering
    \includegraphics[width=\columnwidth]{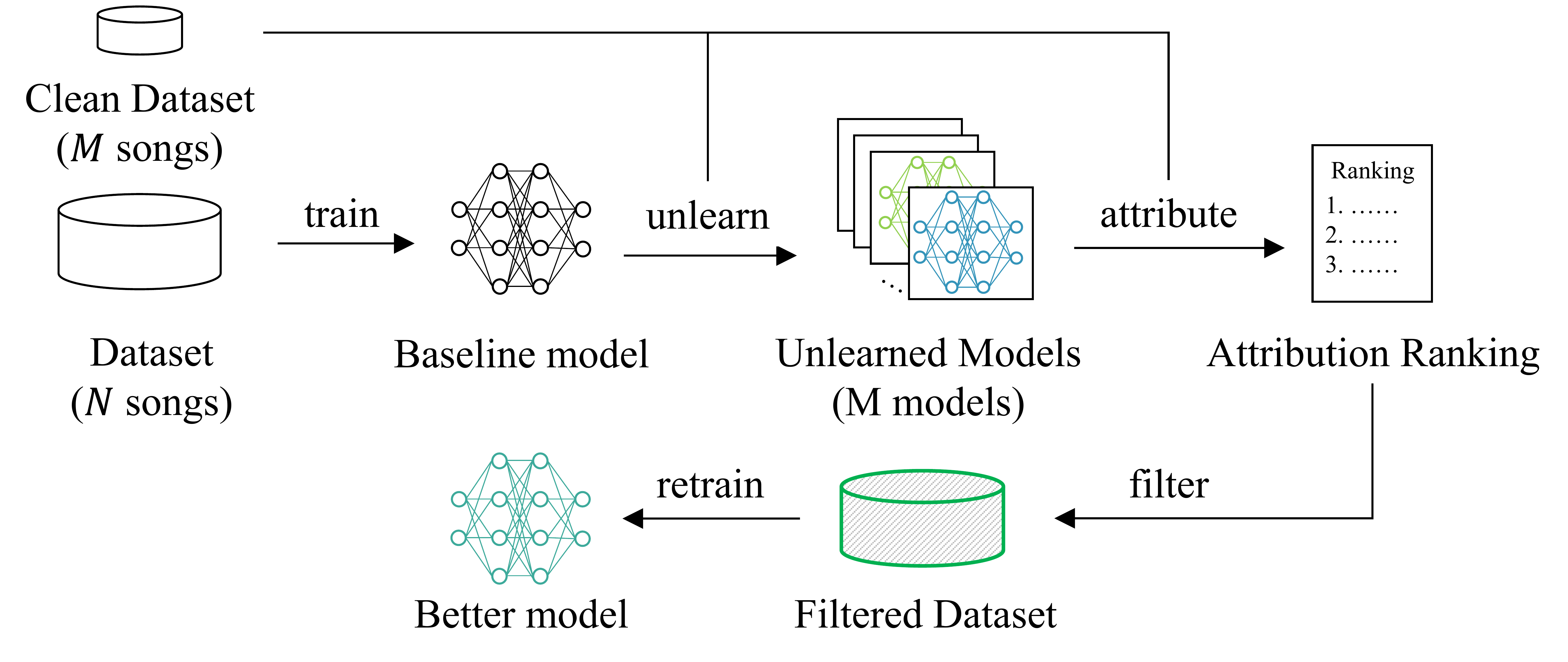}
    \caption{Overview of the unlearning-based data cleaning method.}
    \label{fig:overview-unlearning}
\end{figure}

As shown in Figure~\ref{fig:overview-unlearning}, our unlearning-based data cleaning approach proceeds as follows. First, we train a baseline model on the raw dataset. Next, we unlearn a small set of trusted clean evaluation samples as if they were part of the training data, and compute the average attribution of each training sample with respect to these clean samples.  Training samples with the lowest attribution scores are considered for removal, and a specified ratio of them is filtered out. Filtering can be done in either a unified or per-target manner. Finally, the model is retrained on the remaining samples.

\vspace{1mm}\noindent\textbf{Attribution by Unlearning ---}
Following \cite{wang_data_2025}, we update the model parameters to maximize the loss on unlearned samples in order to forget them. A simple mathematical form can be written as
\begin{equation*}
    \theta' \leftarrow \theta + \alpha \nabla \mathcal{L}(\vec{x}, \theta),
\end{equation*}
where $\theta$ represents the model weights before an unlearning step, $\theta'$ represents the updated model weights, $\alpha$ is the learning rate, and $\nabla \mathcal{L}(\vec{x}, \theta)$ is the gradient of the loss function with respect to the current weights and the unlearned sample $x$.

However, this naive implementation is prone to catastrophic forgetting, where the model not only forgets the characteristics of a single data point, but also loses generalization ability on the rest of the dataset \cite{zhang_catastrophic_2025, kirkpatrick_overcoming_2017}. To mitigate this, we adopt the regularization strategy of elastic weight consolidation (EWC), which preserves pretraining knowledge through the Fisher Information Matrix (FIM) \cite{kirkpatrick_overcoming_2017,wang_data_2025}. Similar to \cite{wang_data_2025}, the final update rule can be written as
\begin{equation*}
    \theta' \leftarrow \theta + \frac{\alpha}{N} \mathbf{F}^{-1} \nabla \mathcal{L}(\vec{x}, \theta),
\end{equation*}
where $\mathbf{F}$ corresponds to the pre-calculated FIM, and $N$ is the number of original training samples.

The FIM quantifies the importance of each model parameter to the inference task. Parameters with high $\mathbf{F}$ values are crucial, and modifying them would degrade model performance. Thus, EWC reduces changes to these crucial parameters during unlearning to avoid catastrophic forgetting~\cite{wang_data_2025, kirkpatrick_overcoming_2017}. The FIM is computed with a diagonal approximation, which can be written as
\begin{equation*}
\mathbf{F}_{ii} = \frac{1}{N} \sum_{x \in D} \left(\frac{\partial \mathcal{L}(\vec{x}, \theta)}{\partial \theta_i}\right)^2 ,
\end{equation*}
where $\mathbf{F}_{ii}$ is the $i$-th element on the diagonal of the FIM, $D$ represents the training dataset, $N = |D|$ is its total number of samples, and $\frac{\partial \mathcal{L}}{\partial \theta_i}$ is the gradient of the loss with respect to a single model parameter $\theta_i$.

To compute data attribution using this formulation, we first compute the reference training loss for each training song $x_i$ ($i=1, \dots, N$) on the baseline model $\theta$, which produces $N$ baseline losses, each denoted as $\mathcal{L}_i$. Then, we unlearn a set of $M$ clean samples $y_j$ ($j=1, \dots, M$), forming $M$ unlearned models, each denoted as $\theta'_j$. Finally, we compute the attribution by measuring the change in loss of each training sample $x_i$ after unlearning, by evaluating $x_i$ on $\theta'_j$, which gives us a total of $N \times M$ losses $\mathcal{L}'_{i,j}$. We subtract the final loss from the initial loss to measure the loss change $\Delta\mathcal{L}_{i,j} = \mathcal{L}'_{i,j} - \mathcal{L}_i$.
For architectures such as Open-Unmix~\cite{stoter_open-unmix_2019}, where each target is an independent model, the procedure above is repeated for each target $t \in \{\text{vocals}, \text{bass}, \text{drums}, \text{other}\}$.

\vspace{1mm}\noindent\textbf{Unified vs Per-target Filtering ---}
After computing the attribution scores $\Delta\mathcal{L}$ for each training sample relative to each clean sample (and target), we can filter the dataset in two ways: a \textit{unified} approach, which uses a single ranking for all stems, or a \textit{per-target} approach, which creates an independent cleaned dataset for each target instrument model.
In the \textit{unified} approach, we calculate a single attribution score for each training song by averaging its $\Delta\mathcal{L}$ across all clean samples and targets:
\begin{equation*}
\Delta\mathcal{L}_i = \frac{1}{M} \sum_j \Delta\mathcal{L}_{i, j}.
\end{equation*}
All songs are ranked by this score. Songs at the bottom $1-r$ are removed, where $r\in(0,1]$ denotes the proportion of data retained after cleaning. The resulting dataset is used to retrain all target models.

In the \textit{per-target} approach, we rank training samples separately for each target $t$ based on their mean attribution for that target:
\begin{equation*}
\Delta\mathcal{L}_i^{(t)} = \frac{1}{M_t} \sum_{j\in t} \Delta\mathcal{L}_{i, j} ,
\end{equation*}
where $M_t$ is the number of clean tracks corresponding to target $t$. Then, we create an independent cleaned dataset for each target by removing the bottom $1-r$ proportion of the samples with the lowest attribution scores for that target. Each target model is then retrained on its respective cleaned dataset.

\subsection{Data Cleaning using a Distributional Metric}

We also propose a data cleaning method based on distributional metrics.
Unlike sample-level metrics such as mean squared error, distributional metrics compare across sample populations, allowing us to quantify the similarity more accurately between datasets of different sizes.
In this work, we use the \Frechet{} Audio Distance (FAD) \cite{kilgour_frechet_2019, gui_adapting_2024}, a widely used distributional metric for evaluating audio quality. FAD leverages a pretrained model to capture high-level audio features.
Inspired by its effectiveness in assessing audio quality, we propose a straightforward filtering method using FAD to identify and remove noisy samples from a training dataset.

Our FAD-based data cleaning method begins by quantifying the perceptual dissimilarity of each training song to a small and clean reference set. Using the per-song FAD method detailed in \cite{gui_adapting_2024}, we compare the set of embedding time-frames from one song to all time-frames across all songs in the reference set.
We use MERT \cite{li_mert_2023} and CLAP \cite{elizalde_natural_2024} as the underlying embeddings for this calculation since they were found to be the most effective in capturing musical differences \cite{gui_adapting_2024}. After computing a score for each song, we use the same approach to rank, filter, and retrain as detailed in the unified unlearning approach.

\section{Evaluation}

\subsection{Experiment Setup}

To evaluate our two data cleaning methods, we created a semi-synthetic dataset named ``Mixed23'' to simulate contamination with multiple types of errors found in real recordings. To ensure the performance gap was clearly observable, the prevalence and intensity of artifacts were made more pronounced than what is typical in studio recordings. The dataset comprises $N = 200$ songs: 100 clean samples from the MUSDB18-Training set \cite{rafii_musdb18_2017}, 50 samples with label noise from SDXDB23\_LabelNoise~\cite{fabbro_sound_2024}, and 50 samples with audio bleeding from SDXDB23\_Bleeding~\cite{fabbro_sound_2024}. The $M = 50$ samples from MUSDB18-Test is used as the clean reference in unlearning, FAD, and MLP.

The Open-Unmix model architecture~\cite{stoter_open-unmix_2019} was selected for experimentation due to its simplicity and training efficiency.
To ensure a robust evaluation of our methods, we trained each model with three different seeds for 500~epochs without early stopping, and picked the epoch with the best validation loss on the MUSDB18-Test set. We evaluated them with the SDR metric~\cite{vincent2006performance} on the hidden evaluation dataset of the Music Demixing Challenge 2021, which consists of 27 out of 30 tracks in MDXDB21~\cite{mitsufuji_music_2022}. We take the median SDR across test songs and report the average performance across three seeds.

\subsection{Baseline Cleaning Approaches}

We compare our proposed methods with several baselines, including Open-Unmix models trained on the contaminated Mixed23 dataset and the original MUSDB18~\cite{rafii_musdb18_2017} training dataset containing 100 clean samples. In addition, inspired by Koo et al.~\cite{koo_self-refining_2023}, we developed a classifier-based baseline method leveraging prior knowledge about label noise and bleeding. Specifically, we use a 4-class multi-layer perceptron (MLP) instrument classification model trained on 50 clean songs from MUSDB18-Test. The MLP model is designed to classify individual 768-dimensional audio frame embeddings extracted using the MERT-v1-95M model~\cite{li_mert_2023}. The MLP has an input layer that projects these embeddings into a 256-dimensional hidden space, followed by a single hidden layer with a ReLU activation and a dropout rate of 0.5. A final linear layer maps the hidden representation to one of the four output classes: vocals, bass, drums, or other. The model then predicts and averages the class probabilities over all frames of a given song. Songs are then sorted based on the average predicted probability for their correct instrument class (i.e., for a vocal stem, the probability of the `vocals' class). Similar to our unlearning filtering method, a percentage of the songs with the lowest probability is removed. We experimented with different ratios~$r$ for this baseline MLP approach.

\section{Results}

\subsection{Finding the Optimal Ratio}

\begin{figure}[t]
    \centering
    \includegraphics[width=0.97\linewidth]{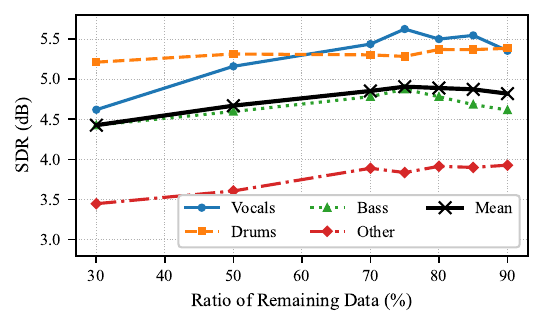}
    \vspace{-15pt}
    \caption{SDR vs.\ data cleaning ratio for unified unlearning via Open-Unmix on the Mixed23 dataset.
    }
    \label{fig:ratios-unified}
\end{figure}

In our experiments, we discovered that the ratio~$r$ for filtering out low-attribution samples is dependent on the specific characteristics of the dataset and the model architecture. Therefore, independently for each considered method, we performed the filtering process with multiple ratios, ranging from removing 5\% to 50\% of the data.
The results are shown in Figure~\ref{fig:ratios-unified}. The best $r$ for cleaning the Mixed23 dataset on the Open-Unmix model is determined to be $r=0.75$ (where 25\% of data is removed), achieving a mean SDR of 4.91\,dB.
We applied the same method to find the optimal ratio for FAD and MLP as well, with $r=0.5$ found to be optimal for both approaches.

\subsection{Result Comparison}
\begin{table*}[t]
    \centering
    \footnotesize
    \vspace{-6pt}
    \caption{SDR scores of different training datasets on the Open-Unmix model. Each SDR score is the mean of three seeds.}
    \label{tab:performance}
    \vspace{2mm}
    \begin{tabular}{l c c c c c c c c}
        \toprule
        \multirow{2}{*}{\textbf{Training Dataset}} & \multirow{2}{*}{\textbf{Cleaning Method}} & \multirow{2}{*}{\textbf{\begin{tabular}[c]{@{}c@{}}Extra \\ Pretrained Model\end{tabular}}} & \multirow{2}{*}{\textbf{\begin{tabular}[c]{@{}c@{}}\% \\ clean\end{tabular}}} & \multicolumn{5}{c}{\textbf{SDR median [dB] on MDX21 Hidden}} \\ & & & & 
        \textit{vocals} & \textit{bass} & \textit{drums} & \textit{other} & \textbf{\textit{avg}} \\
        
        \midrule
        
        Mixed23 & No cleansing & - & 50\% & 5.52 & 4.67 & 5.49 & 3.70 & 4.85 \\
        MUSDB18-Train (clean) & - & - & 100\% & 5.40 & 4.87 & 5.72 & 3.77 & 4.94 \\
        MUSDB18-Test (clean) & - & - & 100\% & 5.40 & 4.42 & 4.95 & 3.66 & 4.61 \\
        Mixed23 & MLP, MERT, $r=0.5$ & MERT-v1-95M & 77\% & 5.63 & 5.02 & 5.56 & 3.78 & 5.00 \\
        
        \cdashline{1-9} \rule{-2pt}{2.5ex}
        
        Mixed23 & Unlearning, targets, $r=0.75$ & - & 62\% & 5.66 & 4.88 & 5.28 & 3.83 & 4.91 \\
        Mixed23 & Unlearning, unified, $r=0.75$ & - & 61\% & 5.62 & 4.88 & 5.28 & 3.84 & 4.91 \\
        Mixed23 & FAD, MERT, $r=0.5$ & MERT-v1-95M & 70\% & 5.53 & 4.91 & 5.41 & 3.74 & 4.90 \\
        Mixed23 & FAD, CLAP, $r=0.5$ & CLAP-2023 & 72\% & 5.43 & 4.88 & 5.65 & 3.70 & 4.91 \\
        
        \bottomrule
    \end{tabular}
\end{table*}
\begin{table}[t]
  \centering
  \footnotesize
  \vspace{-6pt}
  \caption{SDR results for EffectsDB (EDB), designed to test generalization to unseen audio effects.
  These models are trained for 200 epochs instead of 500 epochs as in Table~\ref{tab:performance}.}
  \label{tab:effects_db_condensed}
  \vspace{2mm}
  \resizebox{\columnwidth}{!}{%
  \begin{tabular}{@{}l c c c c c@{}}
    \toprule
    \multirow{2}{*}{\textbf{Dataset \& Method}} & \multicolumn{5}{c}{\textbf{SDR median [dB] on MDX21 Hidden}} \\ 
    & \textit{vocals} & \textit{bass} & \textit{drums} & \textit{other} & \textbf{\textit{avg}} \\
    \midrule
    MUSDB18-Train (clean) & 5.10 & 4.63 & 5.16 & 3.62 & 4.63 \\
    EDB - No cleansing & 4.30 & 4.32 & 5.21 & 3.20 & 4.25 \\
    EDB - MLP, MERT, $r=0.9$ & 4.40 & 4.39 & 5.00 & 3.26 & 4.26 \\
    \cdashline{1-6} \rule{-2pt}{2.5ex}
    EDB - Unlearn, unified, $r=0.9$ & 4.23 & 4.41 & 5.50 & 3.25 & 4.35 \\
    EDB - FAD, CLAP, $r=0.9$ & 4.59 & 5.51 & 4.30 & 3.36 & 4.44 \\
    \bottomrule
  \end{tabular}
  }
\end{table}
We now report on the main results. As shown in Table~\ref{tab:performance}, our proposed data cleaning methods demonstrate a notable improvement in model performance over the model trained with the contaminated Mixed23 data. This model achieves a mean SDR of 4.85\,dB. The model trained only on the small, clean reference set used for unlearning (MUSDB18-Test) performs worse, at 4.61\,dB. In contrast, the upper-bound model trained on the entirely clean MUSDB18-Training dataset with 100 clean songs reaches 4.94\,dB, representing the performance target for our methods.

Both of our unlearning-based methods reach a mean SDR of 4.91\,dB. This result is noteworthy: the cleaned dataset outperforms both the original contaminated dataset (4.85\,dB) and the clean reference set (4.61\,dB) that was used to guide the cleaning process. This demonstrates that our method effectively combines the strengths of both datasets. Furthermore, this enhancement effectively closes two-thirds of the gap between the contaminated baseline and the clean-data baseline ($(4.91 - 4.85)/(4.94 - 4.85) = 0.667$). Similarly, the FAD-based filtering method also improves the score from 4.85 to 4.91\,dB. CLAP performed similarly to MERT as the underlying embedding for FAD filtering. These results validate the effectiveness of our proposed noise-agnostic criteria to identify and remove detrimental samples from a noisy dataset.

It is important to note that the MLP-based filtering method achieves the highest overall score (5.00\,dB). However, the MLP baseline was carefully designed and trained to be effective specifically at detecting label noise and audio bleeding. While it is highly effective for this specific task, such an approach lacks generalizability, which we explicitly test in Sec.˜\ref{subsec:generalization}. Notably, and in contrast to the MLP, our proposed unlearning- and FAD-based methods are fundamentally noise-agnostic. They do not rely on prior knowledge of the specific data defects. This makes these two proposed approaches more robust and broadly applicable to cleaning diverse datasets without requiring noise-specific solutions.

\subsection{Remaining Clean Samples and Model Performance}

To better understand the filtering mechanisms, we analyzed the percentage of clean samples (from the MUSDB18 portion of Mixed23) remaining after applying each cleaning method. In general, we observe that this percentage does not directly correlate with demixing performance (Table~\ref{tab:performance}). Although the dataset cleaned by FAD retained significantly more clean samples than unlearning (72\% vs.~62\%), their resulting models performed similarly (4.91\,dB). More surprisingly, the MLP-based method achieved a final SDR of 5.00\,dB while only retaining 77\% of clean samples, outperforming the 100\% clean MUSDB18-Train baseline (4.94\,dB). These results suggest a complicated reality that some contaminated samples might contain information beneficial to model performance.

As an extra confirmation, we conducted a follow-up experiment by applying the unlearning pipeline to clean MUSDB18-Train. We tested unified and per-target unlearning on a single seed with $r \in \{0.7, 0.9\}$. We achieved 4.95\,dB SDR with per-target unlearning $r = 0.9$, a marginal improvement of 0.01 dB while having 10\% less training data. This indicates that MUSDB18-Train contains no contaminated data and suggests that the improvements we observed in Table~\ref{tab:performance} may have resulted from beneficial contaminated samples.

\subsection{Generalization to Unseen Audio Effects}
\label{subsec:generalization}
To test the capability of the considered methods to generalize to further unknown artifacts, we introduced a new dataset: EffectsDB (EDB). It contains 100~clean songs from MUSDB18-Train and 100~random songs from MoisesDB \cite{pereira_moisesdb_2023}. The latter 100~songs were processed with 3 audio effects: distortion (25 dB), reverb (room size: 0.8, damping: 0.8, wet: 0.5, dry: 0.4), or a low-pass filter (3\,kHz cutoff), applied to random thirds of the set. We ran the experiment for unified unlearning, MLP, and FAD on CLAP, training for 200~epochs on a single seed with $r \in \{0.5, 0.7, 0.9\}$.

As shown in Table~\ref{tab:effects_db_condensed}, all three methods performed the best at $r = 0.9$. Notably, this time, the MLP-based method, which excelled for Mixed23, performs similarly to the dirty baseline on EDB (4.26 vs.\ 4.25\,dB). This suggests that an MLP model designed to remove bleeding and label noise fails to generalize to other audio degradations. In contrast, both the FAD and unlearning methods provide a notable improvement over the dirty baseline (4.44 and 4.35 vs.\ 4.25\,dB), demonstrating their superior ability to generalize in a blind data cleaning scenario.

\subsection{Computational Cost}

We measured all computation on 4 NVIDIA H100 GPUs. The initial baseline model training took 19~hours. The unlearning pipeline took 3~hours, including FIM computation, unlearning, and attribution. The most significant overhead was finding the optimal filtering ratio---we re-trained for 7 different ratios, totaling 98~hours. This brings the entire data cleaning pipeline to a total of 120~hours. In comparison, the FAD approach took 1~hour to compute per-song FAD scores on the CLAP embedding model, and a similar time to re-train. Combined with the initial baseline training and retraining time for 7~ratios~$r$, it took a total of 118~hours.

\section{Conclusion}

In conclusion, our experimental results confirm that the proposed noise-agnostic methods can close 66.7\% of the performance gap for music source separation models trained on contaminated datasets. Crucially, it produced a final dataset that is more effective for training than both the initial contaminated set and the small reference set used to guide the unlearning. Our generalization experiment also showed that noise-agnostic methods are more robust to unseen data corruptions than specialized models. This demonstrates that it is possible to curate higher-quality datasets without needing specialized tools to detect specific artifacts or knowing the exact nature of the noise present in the individual stems.

We must also acknowledge the limitations of our study. The optimal filtering ratio (e.g., removing 25\% of the data) was determined empirically in a resource-intensive process and may vary depending on the dataset's noise level, the model architecture, and the cleaning method. The unlearning attribution method, while effective, also introduces computational overhead compared to a standard training.

For future work, several promising avenues emerge.
First, if the optimal filtering ratio can be determined automatically and without retraining overhead, it would make these techniques more autonomous and easier to apply.
Second, exploring hybrid approaches that combine the broad applicability of a noise-agnostic method with the precision of a specialist classifier could yield even better results.
Third, due to computational resource limitations, we only evaluated our methods on the relatively lightweight Open-Unmix model. Testing these cleaning strategies on larger and more complex architectures, such as HDemucs \cite{defossez_hybrid_2022}, could validate their effectiveness across different model scales.
Finally, the unlearning attribution approach might be generalizable beyond music source separation, and thus applying these data cleaning frameworks to other tasks, such as speech enhancement or sound event detection, could validate their versatility and broader impact.

\newpage

\section{Acknowledgments}

We are grateful for the support of our colleagues. In particular, we wish to thank Giorgio Fabbro for his help with MDXDB21, the hidden evaluation dataset used for Music Demixing Challenge 2021, and Yuichiro Koyama for sharing his insight into optimizing the Open-Unmix training process.

\bibliographystyle{IEEEbib}
\bibliography{refs}

\begin{thebibliography}{10}

\bibitem{zha25data-centric}
Daochen Zha, Zaid~Pervaiz Bhat, Kwei-Herng Lai, Fan Yang, Zhimeng Jiang, Shaochen Zhong, and Xia Hu,
\newblock ``Data-centric artificial intelligence: A survey,''
\newblock {\em ACM Comput. Surv.}, vol. 57, no. 5, Jan. 2025.

\bibitem{fabbro_sound_2024}
Giorgio Fabbro, Stefan Uhlich, Chieh-Hsin Lai, Woosung Choi, Marco Mart{\'\i}nez-Ram{\'\i}rez, Weihsiang Liao, Igor Gadelha, Geraldo Ramos, Eddie Hsu, Hugo Rodrigues, et~al.,
\newblock ``The {S}ound {D}emixing {C}hallenge 2023 - {M}usic {D}emixing {T}rack,''
\newblock {\em arXiv preprint arXiv:2308.06979}, 2023.

\bibitem{kang_improved_2020}
Daniel Kang and Tatsunori Hashimoto,
\newblock ``Improved natural language generation via loss truncation,''
\newblock {\em arXiv preprint arXiv:2004.14589}, 2020.

\bibitem{kim_sound_2023}
Minseok Kim, Jun~Hyung Lee, and Soonyoung Jung,
\newblock ``Sound {Demixing} {Challenge} 2023 - {Music} {Demixing} {Track} {Technical} {Report}: {TFC}-{TDF}-{UNet} v3,'' June 2023.

\bibitem{noauthor_davidliujiafengccom_mdx2023_nodate}
``davidliujiafeng/ccom\_mdx2023 at labelnoise\_train,'' .

\bibitem{goswami_iseparate-sdx_2023}
Nabarun Goswami,
\newblock ``{iSeparate}-{SDX},'' SDX2023 challenge, https://github.com/naba89/iSeparate-SDX.

\bibitem{koo_self-refining_2023}
Junghyun Koo, Yunkee Chae, Chang-Bin Jeon, and Kyogu Lee,
\newblock ``Self-refining of pseudo labels for music source separation with noisy labeled data,''
\newblock in {\em Proc. ISMIR}, 2023.

\bibitem{noauthor_kuielabsdx23_2025}
``kuielab/sdx23,'' SDX2023 challenge, https://github.com/kuielab/sdx23.

\bibitem{ko_mirrored_2024}
Myeongseob Ko, Feiyang Kang, Weiyan Shi, Ming Jin, Zhou Yu, and Ruoxi Jia,
\newblock ``The mirrored influence hypothesis: Efficient data influence estimation by harnessing forward passes,''
\newblock in {\em Proc. CVPR}, 2024, pp. 26276--26285.

\bibitem{wang_data_2025}
Sheng-Yu Wang, Aaron Hertzmann, Alexei Efros, Jun-Yan Zhu, and Richard Zhang,
\newblock ``Data attribution for text-to-image models by unlearning synthesized images,''
\newblock {\em Advances in Neural Information Processing Systems}, vol. 37, pp. 4235--4266, 2024.

\bibitem{kilgour_frechet_2019}
Kevin Kilgour, Mauricio Zuluaga, Dominik Roblek, and Matthew Sharifi,
\newblock ``Frèchet audio distance: A metric for evaluating music enhancement algorithms,''
\newblock in {\em Proc. Interspeech}, 2019.

\bibitem{gui_adapting_2024}
Azalea Gui, Hannes Gamper, Sebastian Braun, and Dimitra Emmanouilidou,
\newblock ``Adapting {Frèchet} audio distance for generative music evaluation,''
\newblock in {\em IEEE International Conference on Acoustics, Speech and Signal Processing (ICASSP)}, 2024, pp. 1331--1335.

\bibitem{choi2025large}
Woosung Choi, Junghyun Koo, Kin~Wai Cheuk, Joan Serr{\`a}, Marco~A Mart{\'\i}nez-Ram{\'\i}rez, Yukara Ikemiya, Naoki Murata, Yuhta Takida, Wei-Hsiang Liao, and Yuki Mitsufuji,
\newblock ``Large-scale training data attribution for music generative models via unlearning,''
\newblock {\em arXiv preprint arXiv:2506.18312}, 2025.

\bibitem{zhang_catastrophic_2025}
Zhiwei Zhang, Fali Wang, Xiaomin Li, Zongyu Wu, Xianfeng Tang, Hui Liu, Qi~He, Wenpeng Yin, and Suhang Wang,
\newblock ``Catastrophic failure of llm unlearning via quantization,''
\newblock in {\em Proc. ICLR}, 2025.

\bibitem{kirkpatrick_overcoming_2017}
James Kirkpatrick, Razvan Pascanu, Neil Rabinowitz, Joel Veness, Guillaume Desjardins, Andrei~A Rusu, Kieran Milan, John Quan, Tiago Ramalho, Agnieszka Grabska-Barwinska, et~al.,
\newblock ``Overcoming catastrophic forgetting in neural networks,''
\newblock {\em Proceedings of the National Academy of Sciences}, vol. 114, no. 13, pp. 3521--3526, 2017.

\bibitem{stoter_open-unmix_2019}
Fabian-Robert St{\"o}ter, Stefan Uhlich, Antoine Liutkus, and Yuki Mitsufuji,
\newblock ``Open-unmix-a reference implementation for music source separation,''
\newblock {\em Journal of Open Source Software}, vol. 4, no. 41, pp. 1667, 2019.

\bibitem{li_mert_2023}
LI~Yizhi, Ruibin Yuan, Ge~Zhang, Yinghao Ma, Xingran Chen, Hanzhi Yin, Chenghao Xiao, Chenghua Lin, Anton Ragni, Emmanouil Benetos, et~al.,
\newblock ``{MERT}: Acoustic music understanding model with large-scale self-supervised training,''
\newblock in {\em Proc. of ICLR}, 2023.

\bibitem{elizalde_natural_2024}
Benjamin Elizalde, Soham Deshmukh, and Huaming Wang,
\newblock ``Natural {Language} {Supervision} for {General}-{Purpose} {Audio} {Representations},'' Feb. 2023,
\newblock arXiv:2309.05767.

\bibitem{rafii_musdb18_2017}
Zafar Rafii, Antoine Liutkus, Fabian-Robert Stöter, Stylianos~Ioannis Mimilakis, and Rachel Bittner,
\newblock ``{MUSDB18-HQ} - {A}n uncompressed version of {MUSDB18},'' Aug. 2019.

\bibitem{vincent2006performance}
Emmanuel Vincent, R{\'e}mi Gribonval, and C{\'e}dric F{\'e}votte,
\newblock ``Performance measurement in blind audio source separation,''
\newblock {\em IEEE Transactions on Audio, Speech, and Language Processing}, vol. 14, no. 4, pp. 1462--1469, 2006.

\bibitem{mitsufuji_music_2022}
Yuki Mitsufuji, Giorgio Fabbro, Stefan Uhlich, Fabian-Robert St{\"o}ter, Alexandre D{\'e}fossez, Minseok Kim, Woosung Choi, Chin-Yun Yu, and Kin-Wai Cheuk,
\newblock ``Music demixing challenge 2021,''
\newblock {\em Frontiers in Signal Processing}, vol. 1, pp. 808395, 2022.

\bibitem{pereira_moisesdb_2023}
Igor Pereira, Felipe Araújo, Filip Korzeniowski, and Richard Vogl,
\newblock ``Moisesdb: {A} dataset for source separation beyond 4-stems,''
\newblock in {\em Proc. ISMIR}, 2023.

\bibitem{defossez_hybrid_2022}
Alexandre D{\'e}fossez,
\newblock ``Hybrid spectrogram and waveform source separation,''
\newblock {\em arXiv preprint arXiv:2111.03600}, 2021.

\end{thebibliography}

\end{document}